\documentclass[letterpaper,journal]{IEEEtran}
\usepackage[utf8]{inputenc}
\usepackage{amsmath,amssymb,amsfonts}
\usepackage{array}
\usepackage{balance}
\usepackage{booktabs}
\usepackage{colortbl}
\usepackage{geometry}
\usepackage{graphicx}
\usepackage{listings}
\usepackage{multirow}
\usepackage[caption=false,font=normalsize,labelfont=sf,textfont=sf]{subfig}
\usepackage{stfloats}
\usepackage{tabularx}
\usepackage{textcomp}
\usepackage{tikz}
\usepackage{url}
\usepackage{verbatim}
\usepackage{xcolor}
\usepackage{pgfplots}
\usetikzlibrary{shapes.geometric,arrows.meta,positioning,fit,backgrounds}
\def\BibTeX{{\rm B\kern-.05em{\sc i\kern-.025em b}\kern-.08em
    T\kern-.1667em\lower.7ex\hbox{E}\kern-.125emX}}
\newcommand{\artifacturl}{\url{https://github.com/wndif/VeriBugBench}}
\pgfplotsset{compat=1.17}

\definecolor{realColor}{RGB}{128, 128, 128}
\definecolor{ourColor}{RGB}{58, 106, 179}
\definecolor{cirColor}{RGB}{218, 118, 56}

\begin{document}
\title{VeriBugBench: An Empirically Grounded Framework for Constructing Verilog RTL Debugging Benchmarks}

\author{
Xiankai Meng,
Kejian Feng,
Xinlin Zhao,
Zhuo Zhang,
Yan Lei,
Xiaoguang Mao,
and Jiang Wu*%
\thanks{
Xiankai Meng and Kejian Feng are with the School of Computer and Information Engineering,
Institute for Artificial Intelligence, Shanghai Polytechnic University,
Shanghai, China.
}
\thanks{
Xinlin Zhao is an Independent Researcher,
Sunnyvale, CA, USA.
}
\thanks{
Zhuo Zhang is with the School of Information Technology and Engineering,
Guangzhou College of Commerce, Guangzhou, China.
}
\thanks{
Yan Lei is with Chongqing University,
Chongqing, China.
}
\thanks{
Xiaoguang Mao is with the National University of Defense Technology,
Changsha, China.
}
\thanks{
Jiang Wu is with the Academy of Military Sciences,
Beijing, China (corresponding author, e-mail: wujadeon@outlook.com).
}
\thanks{
This work was supported by the National Natural Science Foundation of China
under Grant 62504255 and the Natural Science Foundation of Shanghai under Grant
25ZR1402173.
}
}

\markboth{VeriBugBench: Verilog RTL Debugging Benchmark Construction}%
{VeriBugBench: Verilog RTL Debugging Benchmark Construction}

\maketitle

\begin{center}
\begin{minipage}{0.94\columnwidth}
\centering\footnotesize\itshape
This work has been submitted to the IEEE for possible publication. Copyright
may be transferred without notice, after which this version may no longer be
accessible.
\end{minipage}
\end{center}

\begin{abstract}
RTL source-level debugging research requires benchmark artifacts that provide
faulty designs together with precise change locations, executable test
stimuli, and reproducible configurations. Available Verilog resources usually
provide only a subset of these elements. We present \textbf{VeriBugBench}, a
framework for constructing Verilog RTL debugging benchmarks through
empirically grounded fault construction, LLM-based testbench enhancement, and
execution-based retention. The mutation library maps recurring,
multi-granularity repair patterns observed in RTL bug-fix histories to 19
executable inverse operators. For each project, an
LLM generates a
design-specific stimulus phase from the clean DUT and original testbench; the
phase is composed with the original testbench for candidate execution.
Applying the framework to 45 open-source projects yields
\textbf{VeriBugBench-v1.0}, with 2,608 executable single-fault instances whose
effects are observable at design outputs. Across the 45 projects, the
assembled testbenches increase mean project-level fault observability from
36.01\% to 39.54\% and improve line coverage and execution-trace diversity on
average. VeriBugBench provides versioned RTL
variants, source-level ground truth, testbenches, and execution artifacts for
evaluating RTL debugging methods.
\end{abstract}

\begin{IEEEkeywords}
RTL debugging,
Verilog,
benchmark construction,
mutation testing,
testbench augmentation
\end{IEEEkeywords}

\section{Introduction}

As hardware design complexity grows, verification effort grows even faster
~\cite{wu2024survey}, and debugging remains a major verification bottleneck
~\cite{foster2024wilson}. Verilog RTL executes under concurrent and temporal
semantics~\cite{chen2023essence}.
Because fault evidence evolves during execution, source-level debugging must
connect an observed failure to the responsible RTL construct and its time-aware
execution context~\cite{wu2022tarsel}.

Automated support for this process includes dynamic and mutation-based
fault localization~\cite{wu2022detraque,wu2024kummel}, neural and
LLM-assisted RTL debugging~\cite{heidari2025localizing,xu2024meic,
yao2024locationiskey}, and automated RTL repair~\cite{ahmad2022cirfix}.
Although these methods address different stages of debugging, their
development and evaluation depend on reproducible faulty designs, precise
source-level ground truth, test stimuli, and executable project environments.
A debugging benchmark must therefore provide more than isolated faulty files.

Existing resources cover important parts of this need. CIRFix supplies
executable RTL repair cases~\cite{ahmad2022cirfix}, Hardware BugBase curates
reproducible FPGA bugs and debugging infrastructure~\cite{ma2022debugging},
and RTLLM provides clean designs and test environments for RTL-generation
evaluation~\cite{lu2023rtllm}. Curated real-bug artifacts offer direct
provenance but are costly to expand, while mutation can construct faults at
scale only if its operators and execution conditions are credible. Empirical
RTL bug-fix histories can ground those operators~\cite{wu2023mantra,meng2025rtl},
but the resulting candidates must still be activated and exposed by a project
testbench before they become useful debugging instances.

We present \textbf{VeriBugBench}, a framework for constructing reproducible
Verilog RTL source-level debugging benchmarks. It combines empirically
grounded fault injection, project-level testbench enhancement, and
execution-based instance retention. We use the framework to construct
\textbf{VeriBugBench-v1.0}, whose instances package observable single-fault
RTL variants with source-level ground truth and replayable execution
artifacts. The present implementation supports the Verilog accepted by the
Pyverilog front end~\cite{takamaeda2015pyverilog} and the adopted simulation
flow.

The main contributions of this work are:

\begin{itemize}
  \item \textbf{Construction Methodology.}
  The framework maps multi-granularity repair patterns from RTL bug-fix
  histories to executable mutation operators, applies them at structurally
  valid source locations, enhances each project's testbench, and retains
  candidates whose behavioral effects are observable during execution.

  \item \textbf{Benchmark Artifact.}
  We construct VeriBugBench-v1.0 from 45 open-source projects, obtaining
  2,608 executable single-fault RTL instances whose effects are observable
  at design outputs. The artifact includes clean and mutated RTL, original and
  assembled testbenches, operator and source-location ground truth, and
  reproduction configurations and scripts.

  \item \textbf{Empirical Characterization.}
  We derive 19 executable operators from repair-pattern evidence obtained
  from 1,298 retained bug-fix-related commit records across 300 Verilog projects and evaluate
  the constructed benchmark in terms of distributional consistency, fault
  observability, and execution-artifact richness.
\end{itemize}
\section{Background and Related Work}

Verilog RTL combines concurrency, clocked state, and event-driven
scheduling~\cite{chen2023essence}. \mbox{Consequently,} a defect in a
sequential design may require a temporal activation sequence and several cycles
of propagation before becoming visible at a monitored output
~\cite{tang2026pecker,wu2024tartan}. A useful benchmark must therefore pair
faulty source and location ground truth with the activating testbench and
execution environment. Existing methods consume different parts of this
evidence: LLM-based source analyses operate directly on RTL code context
~\cite{yao2024locationiskey,xu2024meic}, whereas semiformal, dynamic, and
spectrum-based methods use counterexample or simulation traces, coverage,
passing/failing outcomes, or learned hit-statement classifications
~\cite{kumar2022aries,wu2022detraque,wu2022tarsel,jones2005tarantula,
wu2024tartan,heidari2025localizing}.
LLM-assisted hardware verification has explored design-specific test-stimulus
generation~\cite{ma2024verilogreader,zhang2025llm4dv}. In VeriBugBench,
LLM-generated stimuli are incorporated at the project level to enhance the
shared test environment used for candidate execution.
Wit-HW generates additional passing witness tests from a bug-triggering case to
refine spectrum-based hardware bug localization~\cite{ma2024withw}.

Existing executable resources serve complementary purposes. CIRFix provides 32
expert-transplanted repair scenarios with faulty RTL, instrumented testbenches,
correctness information, and reproducibility configurations
~\cite{ahmad2022cirfix}; Hardware BugBase provides a testbed of
reproducible bugs and associated debugging tools for open-source FPGA designs
~\cite{ma2022debugging}. BugGen uses
a self-correcting multi-agent LLM pipeline to generate, insert, and validate
functional RTL faults~\cite{jasper2025buggen}. RTLLM, VerilogEval, and RTL-Repo
benchmark RTL generation at task and repository scales, whereas MetRex targets
post-synthesis metric reasoning
~\cite{lu2023rtllm,liu2023verilogeval,allam2024rtlrepo,
abdelatty2025metrex}. VeriBugBench differs by deriving executable mutation
operators from empirical repair-pattern evidence and constructing a multi-project
population of output-observable single faults with explicit operator and location
provenance, project-level testbenches, and replayable execution artifacts.

This article substantially extends our DAC 2023 conference paper,
MANTRA~\cite{wu2023mantra}, which used manually specified, real-bug-grounded
operators for mutation testing and mutant generation. VeriBugBench instead maps
empirical multi-granularity repair patterns~\cite{meng2025rtl} to executable
inverse operators and combines them with structure-driven candidate discovery,
project-level testbench enhancement, and execution-based filtering to construct
a reproducible multi-project RTL debugging benchmark.

\section{VeriBugBench Framework Overview}

\begin{figure*}[!t]
\centering
\resizebox{\textwidth}{!}{%
\begin{tikzpicture}[
  x=1cm, y=1cm, >=Latex, font=\scriptsize,
  box/.style={draw=black!68, fill=gray!5, rounded corners=1.5pt,
              align=center, text width=2.05cm, minimum height=0.84cm,
              inner xsep=3pt, inner ysep=2pt, line width=0.55pt},
  empirical/.style={box, draw=blue!65!black, fill=blue!6},
  construct/.style={box, draw=teal!65!black, fill=teal!6},
  stimulus/.style={box, draw=orange!75!black, fill=orange!8},
  execution/.style={box, draw=violet!65!black, fill=violet!7,
                    text width=2.48cm, minimum height=1.16cm},
  filter/.style={box, draw=violet!65!black, fill=violet!7,
                 text width=2.30cm, minimum height=1.16cm},
  output/.style={box, draw=black!72, fill=gray!10,
                 text width=2.60cm, minimum height=1.55cm},
  flow/.style={-{Latex[length=2mm,width=1.2mm]},
               line width=0.62pt, draw=black!72},
  dependency/.style={-{Latex[length=1.8mm,width=1.1mm]},
                     dashed, line width=0.52pt},
  rowlabel/.style={font=\scriptsize\bfseries, anchor=west}
]
\node[rowlabel, text=blue!65!black] at (-0.10,3.62)
  {Empirical fault-model grounding};
\node[box]       (history)  at (0,2.95) {RTL bug--fix\\histories};
\node[empirical] (patterns) at (2.60,2.95) {Hierarchical,\\multi-granularity\\repair patterns};
\node[empirical, text width=2.35cm] (operators) at (5.30,2.95)
  {Mapping and inversion\\into 19 guarded\\mutation operators};
\draw[flow] (history) -- (patterns);
\draw[flow] (patterns) -- (operators);

\node[rowlabel, text=teal!65!black] at (-0.10,1.47)
  {Structure-driven fault construction};
\node[box]       (rtl)           at (0,0.80) {Clean target\\RTL project};
\node[construct] (analysis)      at (2.60,0.80) {Structural-context\\matching};
\node[construct] (opportunities) at (5.30,0.80) {Legal operator--location\\opportunities};
\node[construct] (candidates)    at (8.00,0.80) {Single-fault\\RTL candidates};
\draw[flow] (rtl) -- (analysis);
\draw[flow] (analysis) -- (opportunities);
\draw[flow] (opportunities) -- (candidates);
\draw[dependency, draw=blue!65!black]
  (operators.south) -- (opportunities.north);

\node[rowlabel, text=orange!75!black] at (-0.10,-0.68)
  {Project-level testbench enhancement};
\node[box]      (original) at (0,-1.35) {Original\\testbench};
\node[stimulus] (prompt)   at (2.60,-1.35) {Candidate-independent\\project-level prompt};
\node[stimulus] (generate) at (5.30,-1.35) {LLM-generated\\stimulus phase};
\node[stimulus, text width=2.25cm] (assembled) at (8.00,-1.35)
  {Sanitization and\\phase-based assembly\\with original testbench};
\draw[flow] (original) -- (prompt);
\draw[flow] (prompt) -- (generate);
\draw[flow] (generate) -- (assembled);
\draw[dependency, draw=orange!75!black]
  (rtl.south east) to[out=-35,in=145] (prompt.north west);
\draw[dependency, draw=orange!75!black]
  (original.south) -- ++(0,-0.58) -| (assembled.south);

\node[execution] (execute) at (10.95,0.00)
  {Common execution\\Clean RTL and each candidate\\Same assembled testbench};
\node[filter] (retain) at (13.95,0.00)
  {Execution-based filtering\\Successful run\\Output divergence};
\node[output] (benchmark) at (17.10,0.00) {\textbf{VeriBugBench-v1.0}\\[0.6mm]
  Single-fault RTL variants\\Source-level ground truth\\Original/assembled testbenches\\Manifests and configurations};

\draw[flow, draw=teal!65!black]
  (candidates.east) to[out=0,in=150] (execute.north west);
\draw[flow, draw=orange!75!black]
  (assembled.east) to[out=0,in=-150] (execute.south west);
\draw[flow] (execute) -- (retain);
\draw[flow] (retain) -- (benchmark);
\end{tikzpicture}%
}
\caption{Overall architecture of VeriBugBench. Empirical repair patterns ground
guarded inverse mutation operators, while target-project analysis discovers
legal injection opportunities and clean-DUT-based enhancement produces a
project-level assembled testbench. Candidate faults are retained only when they
execute successfully and produce output divergence from the clean design under
the common testbench.}
\label{fig:veribugbench_overview}
\end{figure*}
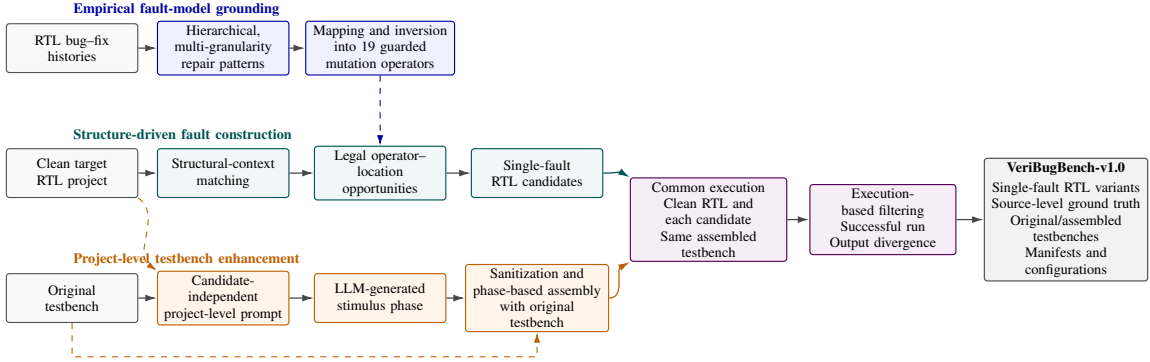

We use the VeriBugBench framework to construct VeriBugBench-v1.0 through the
three coordinated streams shown in Fig.~\ref{fig:veribugbench_overview}.
Empirical repair patterns define executable inverse mutation operators;
structural analysis of each target project identifies valid operator--location
pairs; and the clean DUT with its original testbench provides the context for
project-level testbench enhancement. Candidate RTL variants are compiled and
simulated under the assembled testbench, after which executable faults with
observable output divergence are retained and packaged with source-level
ground truth, testbenches, manifests, and execution configurations.

\section{Empirically Grounded Mutation Framework via Repair-Pattern Mapping}
\label{sec:mutation}

Historical patches may contain coordinated edits, whereas benchmark
construction requires controlled single-location injections with explicit
ground truth. We therefore use multi-granularity repair patterns and their
structural contexts to derive executable inverse operator schemas, which are
instantiated only at context-valid locations in clean target designs and
subsequently filtered by execution.

\subsection{Multi-Granularity Empirical Repair Patterns}

The empirical study of Meng \emph{et al.}~\cite{meng2025rtl} retained 1,298
bug-fix-related commit records from 300 Verilog projects. Each modified RTL
fragment is parsed into an AST, and its buggy and fixed revisions are compared
to obtain a hierarchical edit script. The analysis projects these scripts onto
two complementary views. The expression-level view records the nearest
affected RTL construct, while the statement/root-level view records its
enclosing statement, declaration, or structural context. The published
taxonomy contains 8,756 expression-level and 12,786 statement/root-level AST
edit-action records. We report the two projections separately because the same
repair may contribute information at both levels.

Let $\mathcal{D}=\{(P_{bug},P_{fix})\}$ denote the paired buggy and fixed RTL
fragments. For each pair, $\mathrm{ASTDiff}(P_{bug},P_{fix})$ produces insert,
delete, update, and move actions organized by the AST hierarchy. Statement/root
actions identify repair-bearing statements, declarations, and structural
constructs; expression-level actions identify the affected expression, access,
assignment, event-control, or direction construct within that context.

We normalize a repair pattern as
\begin{equation}
    r=\langle a,n,c\rangle,
\end{equation}
where $a\in\{\mathrm{update},\mathrm{insert},\mathrm{delete},\mathrm{move}\}$,
$n$ is the affected RTL construct, and $c$ is its structurally relevant
enclosing context. Depending on the repair, $n$ may denote an expression
operator, indexed or ranged access, assignment, event control, or declaration,
while $c$ records the surrounding assignment, conditional, case, process, or
module context. This representation preserves the distinction between the
same edit action applied to different RTL constructs or procedural contexts.

\subsection{Repair-Pattern-to-Operator Mapping}

We aggregate the multi-granularity repair records by their mutation-relevant
RTL targets and contexts. Let $\mathcal{N}_{m}$ denote the empirical categories
represented by the operator library and $\mathcal{F}$ the target-construct
families; the mapping $\phi:\mathcal{N}_{m}\rightarrow\mathcal{F}$ combines
parser-level node labels with the same RTL role. Each operator draws on the AST
granularity that defines its target, while cross-level associations supply
enclosing contexts where required.

Operator formation then proceeds in four steps. We examine the edit classes
within each target-construct family, associate the affected construct with its
statement/root context to form $\langle a,n,c\rangle$, and split a family when
different targets or edit directions require distinct inverse transformations.
Contexts that realize the same mutation concept are consolidated. An operator
is retained when its inverse transformation and applicability conditions can
be implemented as a controlled, single-location change in the Pyverilog AST.

Empirical frequency establishes which construct families recur in the repair
data. Operator selection further requires a stable RTL-level interpretation,
an explicit inverse transformation, and implementable applicability
conditions.

A mutation operator $\mu$ is specified by an executable operator schema $S_{\mu}$:
\begin{equation}
    S_{\mu}=\langle m,g,t\rangle,
\end{equation}
where $m$ is an AST matcher, $g$ is a structural guard, and $t$ is a transformation rule. The matcher identifies a target-node family, the guard constrains the enclosing context, and the transformation defines the injected change. Thus, \textit{EdgeFlip} applies only to an event-control edge in an appropriate \texttt{always} context, whereas \textit{NSubDelete} applies to a nonblocking assignment in a statement position. Repair patterns and context-specific implementations may share one paper-level mutation-operator ID when they realize the same mutation concept; the executable schemas retain the guards needed by individual implementations. For example, \textit{ExprUpdate} represents context-valid updates of expression operators rather than a single literal token replacement, while \textit{NSubInsert} aggregates block-, \texttt{if}-, and \texttt{case}-specific implementations.

The direction of this mapping is important. The corpus records a repair
$P_{bug}\xrightarrow{r}P_{fix}$, whereas injection applies the inverse
$P_{clean}\xrightarrow{\mu=r^{-1}}P_{mut}$. For updates and moves, the observed
transformation is reversed; insertion and deletion exchange roles.

\subsection{Empirical Evidence and Operator Support}

Table~\ref{tab:empirical_analysis_views} organizes the 19 operators into five
target-construct families and reports the associated empirical AST categories.
Expression, access, and assignment constructs account for 90.31\% of
the expression-level taxonomy. The five represented families together account
for 8,307 of 8,756 records (94.87\%); the remaining 449 records concern
\textit{FunctionCall} and \textit{PARAMETRIZATION}, for which the current
library defines no operator. These percentages describe the prevalence of the
selected constructs rather than the distribution of generated mutants.
Statement/root records may directly support statement-level targets or supply
the enclosing assignment, conditional, case, process, declaration, and module
contexts required by operator guards.

\begin{table*}[!t]
\centering
\caption{Empirical construct-family organization of the 19 mutation operators. Operator names denote the inverse injection direction.}
\label{tab:empirical_analysis_views}
\renewcommand{\arraystretch}{1.08}
\setlength{\tabcolsep}{4pt}
\footnotesize
\begin{tabularx}{\textwidth}{
>{\raggedright\arraybackslash}p{0.14\textwidth}
>{\raggedright\arraybackslash}p{0.28\textwidth}
X}
\toprule
\textbf{Family} & \textbf{Empirical AST categories and records} & \textbf{Mutation operators in family} \\
\midrule
Expression & \textit{InfixExpression}, \textit{PrefixExpression}, \textit{TERNARY}: 4,710 (53.79\%) &
ExprUpdate, ExprInsert, ExprDelete \\
\midrule
Bit-vector \& access & \textit{MemberAccessing}, \textit{DOWNTO}: 1,873 (21.39\%) &
PartselectUpdate, PartselectInsert, PartselectDelete; PointerUpdate, PointerInsert, PointerDelete \\
\midrule
Assignment & \textit{Assignment}: 1,325 (15.13\%) &
NSubUpdate, NSubInsert, NSubDelete, NSubMove; AssignN2B, AssignB2N \\
\midrule
Timing & \textit{RISING}, \textit{FALLING}: 287 (3.28\%) &
EdgeFlip, EdgeInsert, EdgeDelete \\
\midrule
Port & \textit{HdlDirection}: 112 (1.28\%) & IOFlip \\
\bottomrule
\end{tabularx}
\end{table*}

Together, the five families yield the 19 operators listed in
Table~\ref{tab:empirical_analysis_views}. The same IDs are used in the
implementation and dataset-level reporting; matcher, guard, and transformation
details are encoded by their executable schemas.

\paragraph{Illustrative directional mapping trace.}
Consider a repair that inserts a missing nonblocking assignment into an \texttt{if} branch. AST differencing records the inserted assignment together with its enclosing conditional context, forming an insertion pattern $\langle a,n,c\rangle$. Because benchmark construction proceeds from fixed to faulty code, this pattern supports the inverse operator \textit{NSubDelete}. Its executable definition matches a nonblocking assignment in the corresponding statement context and removes it from the clean design. The empirical insertion and the injected deletion thus describe opposite directions of the same structural change.

\subsection{Structure-Driven Candidate Discovery and Instance Construction}

Given a target RTL project $P$ and operator library $\mathcal{M}$, the
framework collects the context-valid opportunities $\mathcal{O}(P)$ and applies
a configurable candidate policy $\pi$ under construction budget $B$:
\begin{equation}
\begin{aligned}
\mathcal{O}(P)=\{(\mu,f,v)\mid {}&\mu\in\mathcal{M},\ v\in f,\\[-2pt]
                                 &m_{\mu}(v)\land g_{\mu}(v)\},\\
\mathcal{C}(P)={}&\pi\bigl(\mathcal{O}(P);B\bigr).
\end{aligned}
\end{equation}
Here, $f$ is a source file and $v$ is an AST node. Candidate locations follow
from the target design's structure rather than a predefined fault-location
profile.

The policy selects a subset of the discovered opportunities while preserving
single-location injection and source-level ground truth. The choice of $\pi$
is configurable. Each instantiated
candidate contains one injected transformation and retains its operator name,
source file, AST-node identifier, and source-line metadata.

\subsection{Execution-Based Filtering and Observed Distribution}

We retain $P_{mut}$ only if it compiles and differs behaviorally from
$P_{clean}$ under the associated testbench $\mathcal{TB}$. Compilation removes
transformations that are invalid in the concrete design context, while
behavioral filtering removes equivalent or unobserved mutants. The retained
population therefore contains executable candidates whose effects are
observable under the associated testbench.

The resulting operator distribution emerges from the structure-driven
opportunity space, operator applicability, the selected instantiation policy,
and execution-based filtering. Section~\ref{sec:eval} compares this retained
distribution with the empirical bug-fix corpus.

\section{LLM-Based Testbench Enhancement}
\label{sec:testbench}

The utility of an executable RTL debugging benchmark depends on whether
the associated test environment can expose behavioral differences between
the clean and faulty designs. The project-provided testbenches in our
source repositories were written for different purposes and vary
considerably in stimulus scope. Some exercise only nominal transactions or
repeatedly apply a narrow input pattern. Such tests may execute the design
successfully while leaving many fault effects unobserved.

VeriBugBench therefore augments, rather than replaces, each original
testbench. The LLM receives the clean DUT and its existing testbench and
generates a design-specific stimulus block. After extraction and sanitization,
the generated phase is
composed with the retained original phase to form one project-level
testbench, which is subsequently applied uniformly to all candidate mutants
of that project.

\subsection{Construction Objective and Assessment Metrics}
\label{subsec:metrics}

Let $\mathcal{TB}_0$ denote the original project testbench and
$\mathcal{S}_{\mathrm{LLM}}$ the generated stimulus phase. The assembled
testbench is
\begin{equation}
  \mathcal{TB}^{+}
  = \mathrm{Compose}(\mathcal{S}_{\mathrm{LLM}},\mathcal{TB}_0),
  \label{eq:tb_compose}
\end{equation}
which we call the \emph{assembled testbench}. The desired outcome is not merely
more stimulus code, but more observable behavioral variation under the
same DUT interface and logging mechanism.

We assess this outcome using candidate-fault observability and three
execution-spectrum metrics. Candidate-fault observability records whether a
mutant produces an output trace distinguishable from that of the clean
design and is defined in Section~\ref{subsec:rq3}. For the spectrum metrics, a
simulation is represented by
\begin{equation}
  \mathbf{M}\in\{0,1\}^{L\times T},
  \label{eq:M}
\end{equation}
where $M_{l,t}=1$ if executable RTL line $l$ is active at time step $t$.
Line coverage is
\begin{equation}
  \mathrm{Cov}=\frac{|\{l:\sum_t M_{l,t}>0\}|}{L}\times 100\%.
  \label{eq:cov}
\end{equation}

Let $\mathbf{r}_l$ be row $l$ of $\mathbf{M}$ and
$\mathcal{A}=\{l:\|\mathbf{r}_l\|_1>0\}$ the active RTL lines. If
$m(\mathbf{r})$ is the number of active lines sharing trace
$\mathbf{r}$, the implemented uniqueness score is
\begin{equation}
  S_{\mathrm{unique}}
  =\frac{1}{|\mathcal{A}|}
   \sum_{l\in\mathcal{A}}\mathbb{I}\!\left[m(\mathbf{r}_l)=1\right].
  \label{eq:sunique}
\end{equation}
It is therefore the fraction of covered lines whose temporal activation
trace is not shared by another covered line. Finally, let
$\mathbf{c}_t$ be column $t$ of $\mathbf{M}$ and let $P(s)$ be the
empirical frequency of column state $s$. Temporal state entropy is
\begin{equation}
  H=-\sum_s P(s)\log_2 P(s).
  \label{eq:entropy}
\end{equation}
Together, $\mathrm{Cov}$, $S_{\mathrm{unique}}$, and $H$ describe
structural reach, line-level trace separation, and temporal diversity.
RQ4 evaluates these metrics on the assembled execution artifacts.

\subsection{Clean-DUT-Based Prompt Construction}
\label{subsec:prompt}

The evaluated configuration instantiates one prompt template from each
project's complete clean DUT RTL and original testbench, including its drivers,
clock/reset conventions, and logging mechanism. The same project-level prompt
combines this context with fixed, candidate-independent operator-family guidance
covering expressions, bit selections, control-flow outcomes, event-control
ordering, and assignment timing. It
requests explicit boundary values, all-zero, all-one, alternating-bit and
walking-one patterns, branch-distinguishing transitions, and back-to-back
transactions. The generated phase must contain valid Verilog stimulus, must not
introduce randomization or duplicate logging, must not call
\texttt{\$finish}, and must set \texttt{llm\_go\_signal=0} on completion. Once
sanitized and assembled, the resulting versioned testbench is directly
replayable, and the complete prompt is retained with the framework
implementation.

\subsection{Stimulus Generation, Sanitization, and Integration}
\label{subsec:composition}

The LLM returns a Verilog stimulus block rather than a complete
replacement testbench. The framework extracts Verilog from the response
and applies a lightweight sanitizer that moves declarations to the
beginning of each \texttt{initial} block. This step handles a common
Verilog compatibility problem in generated code without changing the
stimulus intent.

The sanitized block is inserted as the first stimulus phase. A global
\texttt{llm\_go\_signal}, initialized to one, pauses the main original stimulus
while time-zero initialization and background processes, such as clock
generation and trace logging, remain active. The generated phase sets the
signal to zero on completion, releasing the retained original stimulus. The
existing logger is reused so that both
phases contribute to the same execution artifact. A watchdog is injected
to terminate non-ending simulations and close an identified file handle
where applicable.

The handoff does not reinitialize the DUT; the original phase continues from
the state left by the generated phase.

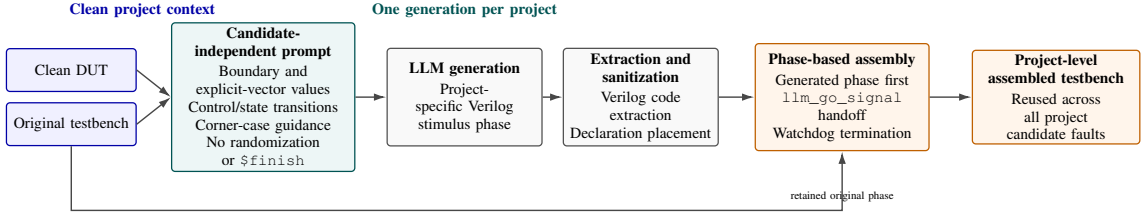
\begin{figure*}[!t]
\centering
\resizebox{\textwidth}{!}{%
\begin{tikzpicture}[
  x=1cm, y=1cm, font=\scriptsize, >=Latex,
  box/.style={rounded corners=1.6pt, align=center, inner xsep=3pt,
              inner ysep=2pt, line width=0.58pt, minimum height=1.22cm},
  input/.style={box, draw=blue!65!black, fill=blue!6,
                text width=1.85cm, minimum height=0.68cm},
  prompt/.style={box, draw=teal!65!black, fill=teal!7,
                 text width=2.70cm, minimum height=1.55cm},
  stage/.style={box, draw=black!68, fill=gray!5,
                text width=2.25cm, minimum height=1.55cm},
  assembly/.style={box, draw=orange!75!black, fill=orange!8,
                   text width=2.55cm, minimum height=1.70cm},
  output/.style={box, draw=orange!75!black, fill=orange!8,
                 text width=2.45cm, minimum height=1.40cm},
  flow/.style={-{Latex[length=2mm,width=1.2mm]},
               line width=0.64pt, draw=black!72}
]
\node[font=\scriptsize\bfseries, text=blue!65!black, anchor=west]
  at (-0.15,1.38) {Clean project context};
\node[input] (dut)      at (0,0.43) {Clean DUT};
\node[input] (original) at (0,-0.43) {Original testbench};

\node[prompt] (prompt) at (3.05,0) {\textbf{Candidate-independent prompt}\\[0.5mm]
  Boundary and explicit-vector values\\Control/state transitions\\
  Corner-case guidance\\No randomization or \texttt{\$finish}};
\node[stage] (generation) at (6.25,0) {\textbf{LLM generation}\\[0.5mm]
  Project-specific Verilog\\stimulus phase};
\node[stage] (sanitize) at (9.05,0) {\textbf{Extraction and sanitization}\\[0.5mm]
  Verilog code extraction\\Declaration placement};
\node[assembly] (assembly) at (12.25,0) {\textbf{Phase-based assembly}\\[0.5mm]
  Generated phase first\\\texttt{llm\_go\_signal} handoff\\Watchdog termination};
\node[output] (result) at (15.65,0) {\textbf{Project-level\\assembled testbench}\\[0.5mm]
  Reused across all project\\candidate faults};

\draw[flow] (dut.east) -- (prompt.west);
\draw[flow] (original.east) -- (prompt.west);
\draw[flow] (prompt) -- (generation);
\draw[flow] (generation) -- (sanitize);
\draw[flow] (sanitize) -- (assembly);
\draw[flow] (assembly) -- (result);
\draw[flow] (original.south) -- ++(0,-1.02) -|
  node[pos=0.73, below, font=\tiny]{retained original phase}
  (assembly.south);
\node[font=\scriptsize\bfseries, text=teal!65!black]
  at (6.25,1.38) {One generation per project};
\end{tikzpicture}
}
\caption{Project-level workflow for LLM-based testbench enhancement. A
candidate-independent prompt constructed from the clean DUT and original
testbench generates a stimulus phase, which is sanitized and composed with the
retained original phase. The assembled testbench is reused across the
project's candidate faults.}
\label{fig:framework}
\end{figure*}

Figure~\ref{fig:framework} summarizes the workflow from prompt construction to
assembled-testbench execution; the resulting artifacts are assessed in RQ3 and
RQ4.

\section{Dataset Construction and Composition}
\label{sec:dataset}

\subsection{Project Sources and Dataset Scope}

To instantiate VeriBugBench-v1.0, we selected 45 open-source Verilog
projects from three sources that differ in origin and source-code scale:

\begin{itemize}
    \item \textbf{Native projects}: 11 open-source projects collected for
    this study. They form the larger-design portion of the selected corpus
    and include hierarchical control and datapath logic.
    
    \item \textbf{CirFix projects}: 11 designs drawn from the executable
    CirFix repair artifact and previously exercised in an RTL repair
    workflow~\cite{ahmad2022cirfix}.
    
    \item \textbf{RTLLM designs}: 23 compact designs originating from an
    RTL-generation benchmark~\cite{lu2023rtllm}. Their reference RTL and project test
    environments serve as clean construction inputs.
\end{itemize}

The selected groups range from compact functional modules to larger
open-source designs. Table~\ref{tab:dataset_comparison} reports their project
sizes and the resulting benchmark scale.

\begin{table*}[!t]
\centering
\caption{Scale and Composition of VeriBugBench-v1.0, with the Original
CirFix Artifact as a Descriptive Reference}
\label{tab:dataset_comparison}
\small
\setlength{\tabcolsep}{4pt}
\begin{tabular}{l c rrrr rrrr}
\toprule
\multirow{2}{*}{\textbf{Source}} & \multirow{2}{*}{\textbf{\# Proj}} & \multicolumn{4}{c}{\textbf{SLOC (NCSL)}} & \multicolumn{4}{c}{\textbf{Faulty RTL Instances}} \\
\cmidrule(lr){3-6} \cmidrule(lr){7-10}
 & & \textbf{Total} & \textbf{Mean} & \textbf{Med.} & \textbf{Min--Max} & \textbf{Total} & \textbf{Mean} & \textbf{Med.} & \textbf{Min--Max} \\
\midrule
\multicolumn{10}{l}{\textit{Original CirFix Artifact (Descriptive Reference)}} \\
CirFix (origin) & 11 & 6,979 & 634.5 & 106.0 & 14--3,305 & 61 & 5.5 & 6.0 & 2--9 \\
\midrule
\multicolumn{10}{l}{\textit{VeriBugBench-v1.0 by Project Source}} \\
CirFix (VeriBugBench) & 11 & 6,979 & 634.5 & 106.0 & 14--3,305 & 810 & 73.6 & 60.0 & 14--186 \\
RTLLM (VeriBugBench) & 23 & 803 & 34.9 & 32.0 & 11--94 & 893 & 38.8 & 33.0 & 7--87 \\
Native (VeriBugBench) & 11 & 22,052 & 2004.7 & 846.0 & 162--12,033 & 905 & 82.3 & 60.0 & 21--160 \\
\midrule
\textbf{Total (VeriBugBench-v1.0)} & 45 & 29,834 & 663.0 & 45.0 & 11--12,033 & 2,608 & 58.0 & 47.0 & 7--186 \\
\bottomrule
\multicolumn{10}{l}{\footnotesize \parbox{0.96\textwidth}{\textit{Note:
The VeriBugBench-v1.0 total aggregates only the three constructed
project-source groups; the original CirFix row is shown as a descriptive
reference and is not included in the total.}}} \\
\end{tabular}
\end{table*}

\subsection{LLM-Based Testbench Preparation}

For each project, the workflow in Section~\ref{sec:testbench} produces a
versioned assembled testbench used during candidate execution and filtering.
Section~\ref{sec:eval} evaluates its
behavioral effect through fault observability and execution-artifact metrics.

\subsection{Mutation Opportunity Analysis}

Applying the operator matchers and structural guards defined in
Section~\ref{sec:mutation} to the 45 project ASTs yields \textbf{167,750}
legal $\langle\text{operator},\text{file},\text{node}\rangle$ opportunities
(Table~\ref{tab:mutation_stats}). Bit-vector and access operators account for
93,957 opportunities, followed by assignment (44,276), expression (24,693),
port (2,640), and timing operators (2,184).

\begin{table*}[!t]
\centering
\caption{Distribution of legal operator--location opportunities across project categories.}
\label{tab:mutation_stats}
\begin{tabular}{lccccccc}
\toprule
{} &      \# Proj. &          Assign. &          Timing &            Expr. &            Port &         Bit-Vec. &             Total \\
Category       &              &                  &                 &                  &                 &                  &                   \\
\midrule
CirFix         &           11 &            6,355 &             212 &            3,645 &             419 &           13,543 &            24,174 \\
Native         &           11 &           37,141 &           1,824 &           20,580 &           2,113 &           78,852 &           140,510 \\
RTLLM          &           23 &              780 &             148 &              468 &             108 &            1,562 &             3,066 \\
\textbf{Total} &  \textbf{45} &  \textbf{44,276} &  \textbf{2,184} &  \textbf{24,693} &  \textbf{2,640} &  \textbf{93,957} &  \textbf{167,750} \\
\bottomrule
\end{tabular}
\end{table*}

These opportunities define the structural input population from which
benchmark candidates are instantiated.

\subsection{Candidate Construction and Execution-Based Filtering}
\label{subsec:execution_filtering}

VeriBugBench-v1.0 samples $b=20$ proposals uniformly with replacement from
each nonempty project--operator opportunity set and collapses repeated
$\langle\text{operator},\text{file},\text{node}\rangle$ identifiers. This
produces \textbf{7,358 instantiated and deduplicated candidates}. The saved
candidate manifests define the reported snapshot and support exact replay.

Each candidate is compiled and simulated with its project-level assembled
testbench. Filtering retains candidates that complete execution and differ
from the clean design in the project-provided CSV observation trace. For each
clean--faulty pair, the comparator loads the traces with their recorded
\texttt{time} field as the index, compares their common row prefix in recorded
row order, and checks every logged observation field as a string. Thus,
\texttt{X}/\texttt{Z} values are treated as observed values, and an instance is
output-observable when at least one compared row differs in any logged field;
trace portions beyond the common prefix do not establish observability. These
checks yield \textbf{2,608 executable and output-observable single-fault
instances}.

The construction funnel is therefore 167,750 legal opportunities, 7,358
instantiated candidates, and 2,608 retained instances. The final population
reflects project structure, proposal sampling, deduplication, executability,
and testbench-conditioned observability.

\subsection{Final Observed Mutation Distribution}

Table~\ref{tab:operator_dist_final} reports the retained counts of the 19
mutation operators defined in Table~\ref{tab:empirical_analysis_views}.
Table~\ref{tab:empirical_analysis_views} organizes the operator library by
construct family, whereas Table~\ref{tab:operator_dist_final} shows the population obtained
after project-specific opportunity discovery, proposal sampling,
deduplication, and execution-based filtering. Section~\ref{sec:eval} compares
this retained distribution with the empirical reference corpus.

\begin{table*}[!t]
\centering
\caption{Observed distribution of retained mutation operators in VeriBugBench-v1.0. Counts and percentages are measured after candidate construction and execution-based filtering. Bar colors encode normalized AST transformation classes: \textcolor{blue!70}{\textbf{update}}, \textcolor{teal!70}{\textbf{insert}}, \textcolor{red!70}{\textbf{delete}}, and \textcolor{violet!70}{\textbf{move}}.}
\label{tab:operator_dist_final}
\renewcommand{\arraystretch}{1.2}
\newcommand{\distbarfinal}[3]{%
    \makebox[6.5em][r]{#1} \textcolor{#2}{\rule{#3}{6pt}}%
}
\begin{tabular}{lll}
\toprule
\textbf{Operator family} & \textbf{Mutation operator} & \textbf{Observed distribution (\% \& count)} \\
\midrule
\multirow{3}{*}{Expression} & ExprUpdate & \distbarfinal{8.13\% (212)}{blue!60}{2.60cm} \\
  & ExprInsert & \distbarfinal{8.40\% (219)}{teal!60}{2.69cm} \\
  & ExprDelete & \distbarfinal{10.93\% (285)}{red!60}{3.50cm} \\
\midrule
\multirow{6}{*}{\shortstack{Bit-vector\\\& access}} & PartselectUpdate & \distbarfinal{2.30\% (60)}{blue!60}{0.74cm} \\
  & PartselectInsert & \distbarfinal{0.31\% (8)}{teal!60}{0.10cm} \\
  & PartselectDelete & \distbarfinal{2.34\% (61)}{red!60}{0.75cm} \\
  & PointerUpdate & \distbarfinal{5.21\% (136)}{blue!60}{1.67cm} \\
  & PointerInsert & \distbarfinal{9.16\% (239)}{teal!60}{2.94cm} \\
  & PointerDelete & \distbarfinal{3.41\% (89)}{red!60}{1.09cm} \\
\midrule
\multirow{6}{*}{Assignment} & NSubUpdate & \distbarfinal{8.28\% (216)}{blue!60}{2.65cm} \\
  & NSubInsert & \distbarfinal{10.74\% (280)}{teal!60}{3.44cm} \\
  & NSubDelete & \distbarfinal{10.01\% (261)}{red!60}{3.21cm} \\
  & NSubMove & \distbarfinal{8.13\% (212)}{violet!60}{2.60cm} \\
  & AssignN2B & \distbarfinal{2.22\% (58)}{blue!60}{0.71cm} \\
  & AssignB2N & \distbarfinal{0.27\% (7)}{blue!60}{0.09cm} \\
\midrule
\multirow{3}{*}{Timing} & EdgeFlip & \distbarfinal{4.83\% (126)}{blue!60}{1.55cm} \\
  & EdgeDelete & \distbarfinal{1.73\% (45)}{red!60}{0.55cm} \\
  & EdgeInsert & \distbarfinal{1.61\% (42)}{teal!60}{0.52cm} \\
\midrule
Port & IOFlip & \distbarfinal{1.99\% (52)}{blue!60}{0.64cm} \\
\bottomrule
\end{tabular}
\end{table*}

\textbf{Dataset availability.} VeriBugBench-v1.0 and its construction
framework are publicly released at \artifacturl{} under the license specified
in the repository. The versioned artifact includes the framework source,
selected project versions and configurations, clean and injected RTL with
operator and source-location ground truth, original and assembled testbenches,
candidate manifests, simulation and coverage metadata, and reproduction
scripts.

\section{Evaluation of VeriBugBench}
\label{sec:eval}

\subsection{Research Questions}

The evaluation aims to answer the following research questions:

\begin{itemize}
    \item \textbf{RQ1 (Benchmark Scale and Composition).} What are the project scale, source composition, and retained-instance scale of VeriBugBench-v1.0?

    \item \textbf{RQ2 (Empirical Distributional Consistency).} To what extent does the retained benchmark preserve the broad edit-location and action tendencies observed in empirical RTL bug-fix data?

    \item \textbf{RQ3 (Fault Observability).} Does LLM-generated testbench augmentation increase the fraction of candidate mutants whose effects are observable at the design outputs?

    \item \textbf{RQ4 (Diagnostic Artifact Richness).} Does the assembled testbench enrich the recorded execution artifacts in terms of structural reach, singleton-line trace uniqueness, and temporal state diversity?

\end{itemize}

\subsection{Experimental Setup}

All experiments were conducted on a Linux server. RTL simulation, mutation execution, and spectrum collection were performed using the unified automation pipeline described in Section~\ref{sec:dataset}.

For each project, we evaluated the project-provided \textbf{Original}
testbench $\mathcal{TB}_0$ and the \textbf{Assembled} testbench
$\mathcal{TB}^{+}$, formed by composing the generated stimulus phase with
$\mathcal{TB}_0$.
All mutation operators, injected faults, and evaluation scripts were held
identical across configurations to ensure fair comparison.

\textbf{LLM generation configuration.} We used GPT-4 with temperature 0.2
and a fixed candidate-independent prompt strategy for each project. Generation
failures were not automatically retried. Because model-service behavior and
LLM generation are stochastic, the released prompts, assembled testbenches,
manifests, and execution configurations define the evaluated benchmark
snapshot; we do not claim that a fresh model call reproduces the same generated
stimulus verbatim.

\subsection{Benchmark Scale and Composition (RQ1)}

Table~\ref{tab:dataset_comparison} summarizes VeriBugBench-v1.0: 45 projects,
29,834 SLOC, and 2,608 executable, output-observable single-fault instances.
Project sizes range from 11 to 12,033 SLOC, and retained instances range from
7 to 186 per project, with a median of 47. The original CirFix artifact is
included only as a descriptive reference because it differs in project scope
and construction procedure.

\textbf{Answer to RQ1:} VeriBugBench-v1.0 combines 45 projects from three
source groups and retains 2,608 executable, output-observable instances,
providing a multi-source and multi-scale RTL debugging benchmark.

\subsection{Empirical Distributional Consistency (RQ2)}
\label{sec:bug_realism}

We analyze the original--mutant RTL pairs of all 2,608 retained instances
using the same AST differencing procedure and taxonomy as the empirical
reference~\cite{meng2025rtl}. A pair may produce multiple AST edit actions;
after projection and within-level deduplication, these actions form analysis
records rather than additional benchmark instances.

The expression-level view assigns each record to the nearest affected AST
category, whereas the statement/root-level view assigns it to the enclosing
statement or structural context. The analysis yields 4,074 expression-level
and 6,446 statement/root-level records. Percentages are normalized independently
within each level,
and these record counts are distinct from the 19-operator instance counts in
Table~\ref{tab:operator_dist_final}.

\begin{figure*}[!t]
    \centering
    \includegraphics[width=\textwidth]{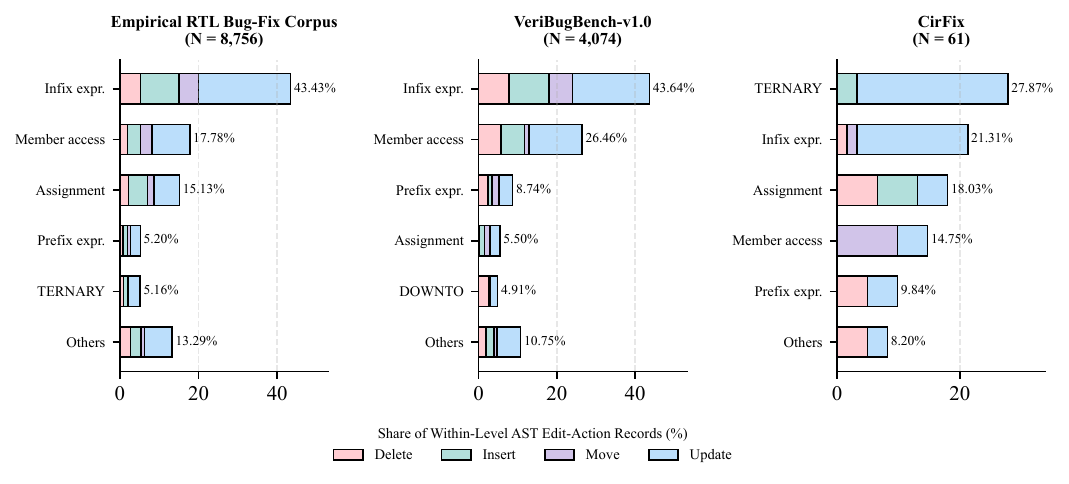}
    \caption{Expression-level distribution of AST edit-action records in the
    empirical RTL bug-fix corpus, VeriBugBench-v1.0, and CirFix. Percentages
    are normalized independently within each source ($N=8{,}756$, $4{,}074$,
    and $61$ records, respectively); colors denote AST edit-action
    classes.}
    \label{fig:expr_comparison}
\end{figure*}

\begin{figure*}[!t]
    \centering
    \includegraphics[width=\textwidth]{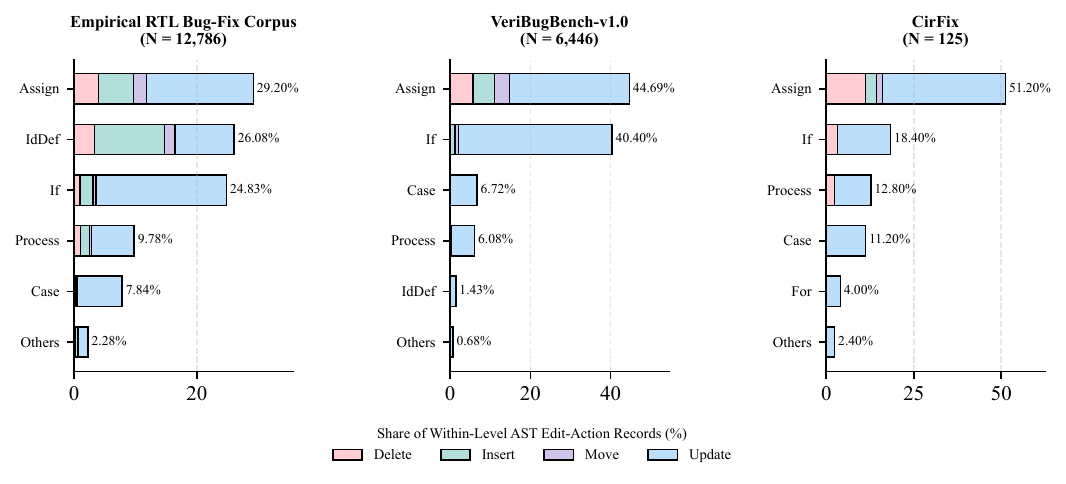}
    \caption{Statement-level distribution of AST edit-action records in the
    empirical RTL bug-fix corpus, VeriBugBench-v1.0, and CirFix. Percentages
    are normalized independently within each source ($N=12{,}786$, $6{,}446$,
    and $125$ records, respectively); colors denote AST edit-action
    classes.}
    \label{fig:stmt_comparison}
\end{figure*}

\textbf{Distribution comparison.} Figures~\ref{fig:expr_comparison} and
\ref{fig:stmt_comparison} compare the expression- and statement-level
distributions with those of the empirical RTL bug-fix corpus and CirFix.

At the expression level, \textit{InfixExpression} and
\textit{MemberAccessing} are the two leading categories in both the empirical
corpus (43.43\% and 17.78\%) and VeriBugBench (43.64\% and 26.46\%). CirFix
shows a different concentration, with \textit{TERNARY} as its largest category
(27.87\%).

At the statement level, VeriBugBench emphasizes \textit{HdlStmAssign}
(44.69\%) and \textit{HdlStmIf} (40.40\%), consistent with their prominence in
the empirical corpus (29.20\% and 24.83\%), but underrepresents
\textit{HdlIdDef} (1.43\% versus 26.08\%). It retains \textit{Case} and
\textit{Process} contexts at 6.72\% and 6.08\%, compared with 7.84\% and
9.78\% empirically. Update remains the largest action class in both
VeriBugBench (47.20\% expression; 82.89\% statement) and the empirical
reference (51.90\%; 64.23\%). The agreement is therefore directional rather
than exact, reflecting the benchmark's focus on executable single-location
transformations.

\textbf{Answer to RQ2:} VeriBugBench preserves broad expression,
statement-context, and edit-action tendencies in empirical RTL bug-fix data,
with differences due to its executable single-location scope. This supports
aggregate empirical grounding, not exact distributional reproduction.

\subsection{Mutation Analysis: Fault Observability (RQ3)}
\label{subsec:rq3}

\textbf{Metric.} For project $p$, let $M_p$ denote the common pool of generated candidate
mutants evaluated under both configurations and
$K_{p,c}$ the number whose behavior is distinguishable from the clean
design under testbench configuration $c\in\{\mathrm{Original},\mathrm{Assembled}\}$.
The project-level mutation score is
\begin{equation}
  \mathrm{MS}_{p,c} \;=\; \frac{K_{p,c}}{M_p} \times 100\%.
  \label{eq:ms}
\end{equation}
A mutant is killed if at least one compared trace row produces an observable
output mismatch under the execution-based criterion in
Section~\ref{subsec:execution_filtering}. Thus, mutation score operationalizes candidate-mutant
observability in this construction study. Because the projects contain substantially different
numbers of candidate mutants, we use the unweighted macro-average
\begin{equation}
  \overline{\mathrm{MS}}_{c}
  \;=\; \frac{1}{|\mathcal{P}|}
  \sum_{p\in\mathcal{P}}\mathrm{MS}_{p,c}
  \label{eq:macro_ms}
\end{equation}
as the primary cross-project measure, so that each project contributes
equally.  The killed-mutant columns in Table~\ref{tab:mutation} are sums
over projects, whereas the mutation-score columns report
$\overline{\mathrm{MS}}_{c}$.  For completeness, we also report the
mutant-level micro-average, $\sum_p K_{p,c}/\sum_p M_p$, in the accompanying
text.

Candidates observable under the assembled testbenches constitute the retained
VeriBugBench-v1.0 population.

\textbf{Results.} Table~\ref{tab:mutation} summarizes the results.  The assembled
testbenches make 352 more candidates observable in aggregate. The mean project-level mutation
score increases from 36.01\% to 39.54\% (+3.53 percentage points), while
the corresponding mutant-level micro-average increases from 30.66\% to
35.44\% (+4.78 percentage points). The largest macro-average gain occurs
in the Native subset (+6.15 percentage points), whereas RTLLM shows a
smaller average gain (+1.77 percentage points). These subset results
describe aggregate tendencies; the augmentation is not guaranteed to
improve every project.

\begin{table*}[!t]
\centering
\caption{Candidate-mutant observability under the original and assembled testbenches.
Observable-mutant counts are summed within each source, whereas MS values are
unweighted means of project-level mutation scores.}
\label{tab:mutation}
\begin{tabular}{lrrrrrrr}
\hline
Source & \#Proj & Total & \multicolumn{2}{c}{Observable} &
         \multicolumn{2}{c}{Mean Project-Level MS (\%)} & Improv. \\
       &        & Mutants & Orig. & Assem. & Orig. & Assem. & (pp) \\
\hline
CirFix & 11 & 1,911 & 708  & 810  & 38.00 & 42.58 & 4.58 \\
Native & 11 & 3,466 & 707  & 905  & 21.34 & 27.49 & 6.15 \\
RTLLM  & 23 & 1,981 & 841  & 893  & 42.08 & 43.84 & 1.77 \\
\hline
Total  & 45 & 7,358 & 2,256 & 2,608 & 36.01 & 39.54 & 3.53 \\
\hline
\end{tabular}
\end{table*}

\textbf{Illustrative project.} The original RAM testbench repeats the same address-zero write/read sequence
100 times, whereas the assembled testbench adds reset handling, address
sweeps, walking-one data, simultaneous read/write operations, and disabled-read
behavior. The number of output-observable candidates increases from 22 of 86
(25.58\%) to 31 of 86 (36.05\%). This project illustrates how design-specific
stimulus broadens the exercised behavior behind the aggregate improvement.

\textbf{Answer to RQ3:} From the common pool of 7,358 candidates, the assembled
testbenches yield 2,608
output-observable instances, compared with 2,256 under the original
testbenches. LLM-generated augmentation therefore increases candidate-mutant
observability on average, although the gain varies across projects and source
subsets.

\subsection{Diagnostic Artifact Richness (RQ4)}
\label{subsec:rq4}

\textbf{Evaluation protocol.} RQ4 computes line coverage
($\mathrm{Cov}$), singleton-line trace uniqueness
($S_\mathrm{unique}$), and temporal state entropy ($H$), defined in
Section~\ref{subsec:metrics}, from the original and assembled execution matrices.
Table~\ref{tab:spectrum} reports unweighted means of project-level values
and paired differences; the Total row averages all 45 projects.

\begin{table*}[!t]
\centering
\caption{Execution-artifact metrics under the original and assembled testbenches. Values and deltas are unweighted means of project-level measurements.}
\label{tab:spectrum}
\begin{tabular}{lrrrrrrrrrr}
\hline
Source & \#Proj &
  \multicolumn{3}{c}{Line Cov.\ (\%)} &
  \multicolumn{3}{c}{$S_\mathrm{unique}$} &
  \multicolumn{3}{c}{Entropy $H$} \\
 & & Orig. & Assem. & $\Delta$ & Orig. & Assem. & $\Delta$ &
       Orig. & Assem. & $\Delta$ \\
\hline
CirFix & 11 & 81.48 & 82.47 & 0.99  & 0.3430 & 0.3725 & 0.0294 & 1.88 & 2.20 & 0.32 \\
Native & 11 & 69.88 & 77.38 & 7.50  & 0.0198 & 0.0372 & 0.0175 & 1.86 & 2.49 & 0.62 \\
RTLLM  & 23 & 71.56 & 73.76 & 2.20  & 0.1954 & 0.2824 & 0.0870 & 1.28 & 1.56 & 0.28 \\
\hline
Total  & 45 & 73.57 & 76.77 & 3.20  & 0.1886 & 0.2445 & 0.0559 & 1.57 & 1.94 & 0.37 \\
\hline
\end{tabular}
\end{table*}

\textbf{Results and answer to RQ4.} Table~\ref{tab:spectrum} shows that
the assembled testbenches increase mean line coverage from 73.57\% to
76.77\% (+3.20 percentage points), $S_\mathrm{unique}$ from 0.1886 to
0.2445 (+0.0559), and temporal state entropy from 1.57 to 1.94 (+0.37).
The latter two metrics can improve even when reachability changes little:
CirFix gains only 0.99 coverage points but increases
$S_\mathrm{unique}$ by 0.0294 and $H$ by 0.32, indicating richer
execution spectra beyond aggregate line coverage. Gains vary across
projects and sources; Native has the largest aggregate coverage and
entropy increases, while RTLLM has the largest uniqueness increase.

\section{Discussion and Threats to Validity}
\label{sec:discussion}

\subsection{Discussion}

VeriBugBench treats benchmark construction as an inspectable pipeline linking
empirical repair patterns, executable mutation operators, structural
opportunities, and execution-based retention. Mapping records, candidate
manifests, and filtering metadata make this provenance auditable. The benchmark
packages paired clean and faulty RTL, operator/location
ground truth, original and assembled testbenches, and execution configurations
for RTL source-level debugging research.

\subsection{Threats to Validity}

\subsubsection{Internal Validity}

Internal validity depends on how empirical repair patterns are mapped to
executable operators and on the sampling, deduplication, tooling, and filtering
configurations. We expose the mappings, operator definitions, candidate
manifests, and execution settings for inspection. The empirical corpus and
benchmark projects overlap slightly. The reported distributional consistency
therefore characterizes this end-to-end construction process. Evaluation
against an independent RTL bug-fix corpus remains future work.

\subsubsection{Construct Validity}

Construct validity is bounded by the adopted measures: aggregate distributional
consistency, candidate-fault observability, and the structural and temporal
properties of execution artifacts. These measures characterize benchmark
construction and artifact quality rather than the performance of a particular
debugging technique.

\subsubsection{External Validity}

VeriBugBench-v1.0 is constructed from 45 open-source Verilog projects spanning
multiple sizes and design categories. Its external validity is therefore
bounded by this project population, the use of single injected source-level
faults, and the Verilog subset accepted by the Pyverilog-based toolchain.
Extending the framework to broader industrial designs, multi-edit or
implementation-dependent faults, and SystemVerilog remains future work.

\subsubsection{Reproducibility}

VeriBugBench-v1.0 supports reproducibility through versioned RTL variants,
original and assembled testbenches, manifests, metadata, execution
configurations and scripts, and documented tool versions. Candidate sampling
and LLM generation are stochastic, so fresh runs may produce different
candidates or stimuli. The released manifests and assembled testbenches
therefore define the benchmark instance evaluated in this paper.

\section{Conclusion}
\label{sec:conclusion}

This paper presented VeriBugBench, a framework for constructing reproducible
Verilog RTL source-level debugging benchmarks, and VeriBugBench-v1.0, a
benchmark constructed from 45 open-source projects. The framework maps
empirical repair patterns to 19 executable mutation operators, discovers
structurally valid mutation opportunities, enhances project-level testbenches
with an LLM, and retains faults through compilation, execution, and
output-divergence checks. VeriBugBench-v1.0 contains 2,608 executable,
output-observable single-fault instances.

The evaluation shows aggregate consistency with empirical edit-location and
repair-action distributions. On average, assembled testbenches improve
candidate-fault observability and all three execution-artifact metrics.
Together, the released benchmark and execution artifacts support reproducible
RTL source-level debugging research.

\section*{Acknowledgment}

\noindent\textit{Generative-AI Disclosure:} OpenAI Codex was used solely for
language editing and iterative refinement of Figs.~1 and~2.

\bibliographystyle{IEEEtran}
\bibliography{reference}

\end{document}